\documentclass{kapproc} 
\usepackage{psfig}
\usepackage{graphicx}

\kluwerbib

\begin{document}

\articletitle{Clusters of Galaxies in X-rays:\\
 Dark Matter}


\author{Africa Castillo Morales}
\affil{Dpto. F\'{\i}sica Te\'orica y del Cosmos,
Universidad de Granada \\                       
Avda. Fuentenueva s/n,                         
18071 Granada, Spain}
\email{acm@ugr.es}

\author{Sabine Schindler}
\affil{Institut f\"ur Astrophysik, Leopold-Franzens-Universit\"at Innsbruck,\\
Technikerstr. 25, A-6020 Innsbruck, Austria}
\email{Sabine.Schindler@uibk.ac.at}

\begin{abstract}
We present the analysis of baryonic and non-baryonic matter distribution in a sample of ten nearby clusters ($0.03<z<0.09$) with temperatures between 4.7 and 9.4 keV. These galaxy clusters have been studied in detail using X-ray data and global physical properties have been determined. Correlations between these quantities have been analysed and compared with the results for distant clusters. We found an interesting dependence between the relative gas extent (expressed as the ratio of gas mass fractions at $r_{500}$ and $0.5\times r_{500}$) and the total cluster mass. The extent of the gas relative to the extent of the dark matter tends to be larger in less massive clusters. This dependence might give us some hints about non-gravitational processes in clusters. The new X-ray satellites Chandra and XMM yield exciting new results for galaxy cluster physics and cosmology. We show the interesting X-ray observations of two intermediate redshift clusters, RBS797 ($z=0.35$) and CL~0939+4713 ($z=0.41$) taken with Chandra and XMM respectively.  

\end{abstract}


\section*{Introduction}

\begin{figure}
\centering
\includegraphics[width=4.9cm]{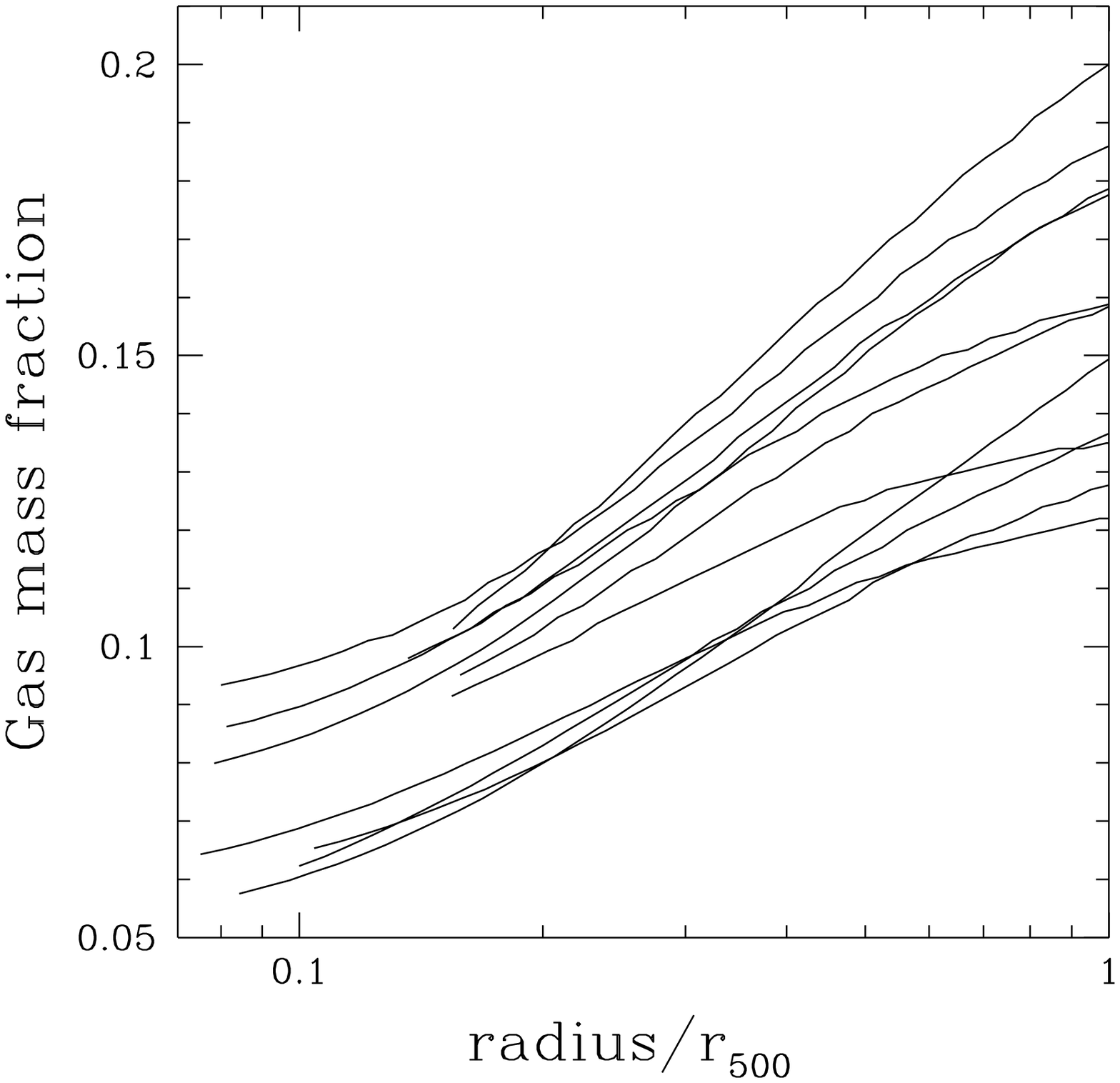}
\includegraphics[width=4.9cm]{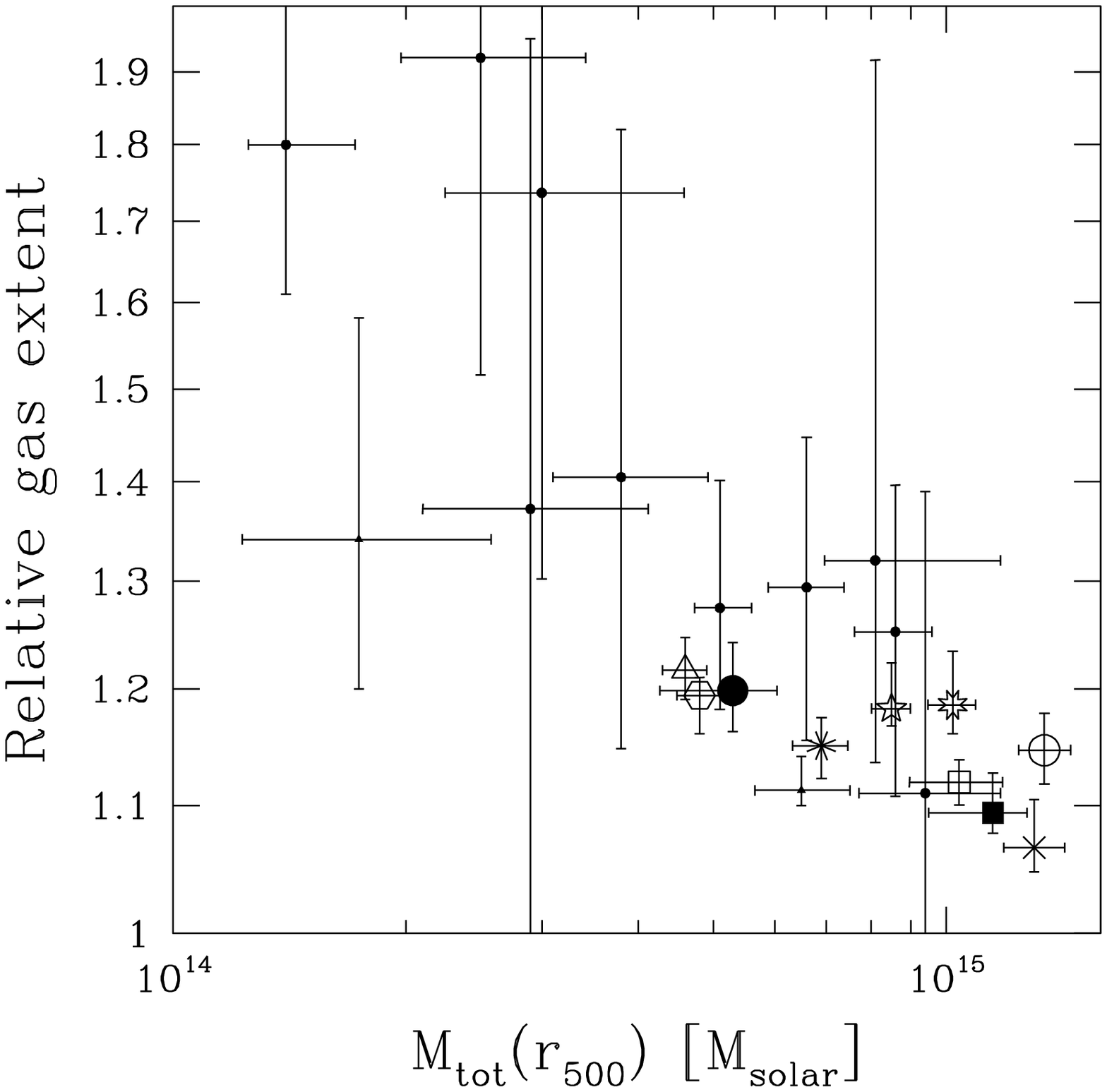}
\caption{Left: Gas mass fraction profiles derived for the nearby cluster sample. Right: Gas extent relative to the dark matter extent expressed as the ratio of gas mass fractions at large radius and small radius (small points represent the distant cluster sample from Schindler 1999 and the larger symbols, the nearby cluster sample by Castillo-Morales \& Schindler in prep.).} 
\label{castillo_fig:fig1}
\end{figure}
The mass fraction of the intra-cluster gas in clusters of galaxies is not negligible. It supposes about 10-30 \% of the total cluster mass. The mass in the galaxies is much smaller, about 3-5 \%. The major fraction of the mass in the cluster is not visible and therefore called dark matter. In this way, measurements of the total mass of the clusters indirectly provide information on the amount and the distribution of the dark matter.

The X-ray emitting gas can be used to trace the cluster potential. With the assumption of hydrostatic equilibrium and spherical symmetry cluster masses can be derived directly from X-ray observations through the gas density gradient, the gas temperature gradient and the gas temperature itself: 
$M(<r)=-\frac{kr}{\mu
m_{p}G}T_{gas}(r)\left (\frac{d\ln \rho_{gas}(r)}{d\ln r}+\frac{d\ln T_{gas}(r)}{d\ln r}\right).$

We selected a sample of ten nearby clusters of galaxies (from the ROSAT PSPC archive) in which an accurate total mass determination is possible. We determine the spatial distributions of gas mass and total mass from X-ray surface brightness and ASCA temperatures (Markevitch et al. 1998) assuming isothermality and using the standard $\beta$-model (Cavaliere $\&$ Fusco-Fermiano 1976) to calculate the gas density profiles. The cosmology used is: $H_{0}$ = 50 km/s/Mpc and $q_{0}$ = 0.5.

\section{Gas mass fraction}

We find that inside each cluster the gas mass fraction increases
outwards (see figure {\ref{castillo_fig:fig1}) implying that the gas distribution is more extended than dark
matter distribution. A low $\Omega$ is required to explain the high gas mass fraction of $<f_{gas}>_{r_{500}}=0.16\pm0.02$, when compared to the baryon fraction predicted by primordial nucleosynthesis. From cluster to cluster we find variations in
the gas mass fraction.
We see no clear trend in the gas mass fraction with the redshift in these data implying an early gas presence in the cluster.

\subsection{Gas extent}

We compare the gas mass fraction at radius $r_{500}$ with the gas mass
fraction at $0.5\times r_{500}$ in each cluster. 
The ratio of these fractions is a measure of how fast the gas mass fraction is increasing with radius and as such a measure for the extent of the gas distribution with respect to the dark mater distribution.
In the nearby sample this relative gas extent E shows a mild
dependence on the total mass (see figure \ref{castillo_fig:fig1}).

This trend can be explained by the physical processes in the gas,
which are assumed to be responsible for the increase of gas mass
fraction with radius like e.g. energy input by supernovae driven
galactic winds. If gas is placed artificially into a model
cluster potential in hydrostatic equilibrium the distributions of gas
and dark matter have the same slope at radii larger than the core
radius, therefore one would expect a priori a ratio $E\approx1$. It
might be that this additional heat input affects low mass clusters
more that massive clusters, so that a massive cluster can maintain a
ratio E = 1 while in the smaller clusters the gas is becoming more extended. Also supportive of energy injections, the work of Ponman et al. 1999 shows that cool ($T<4keV$) clusters observed with ROSAT and GINGA have entropies higher than achievable through gravitational collapse alone, which they explain by pre-heating from strong galactic winds.

\section{Chandra and XMM observations}

In fig.\ref{castillo_fig:fig2} is shown the XMM observation (EPIC MOS 1 \& 2 and PN) of the intermediate distant cluster CL 0939+4713 ($z=0.41$)(De Filippis, Schindler and Castillo-Morales A\&A submitted). The X-ray image shows pronounced substructures. There are two main subclusters which have even some internal structure. This is an indication that the cluster is a dynamically young system. This conclusion is supported by the temperature distribution (see fig.\ref{castillo_fig:fig2}): a hot region is found between the two main subclusters indicating that the cluster is in the process of a major merger, in which the two subclusters will probably collide in a few hundreds Myr. The intra-cluster gas of CL 0939-4713 shows variations of the metal abundances having the optically richer subcluster a somewhat higher metallicity. The detection of such a metallicity variations gives hints on the metal enrichment processes.
\begin{figure}[ht]
\centering
\includegraphics[width=4.8cm]{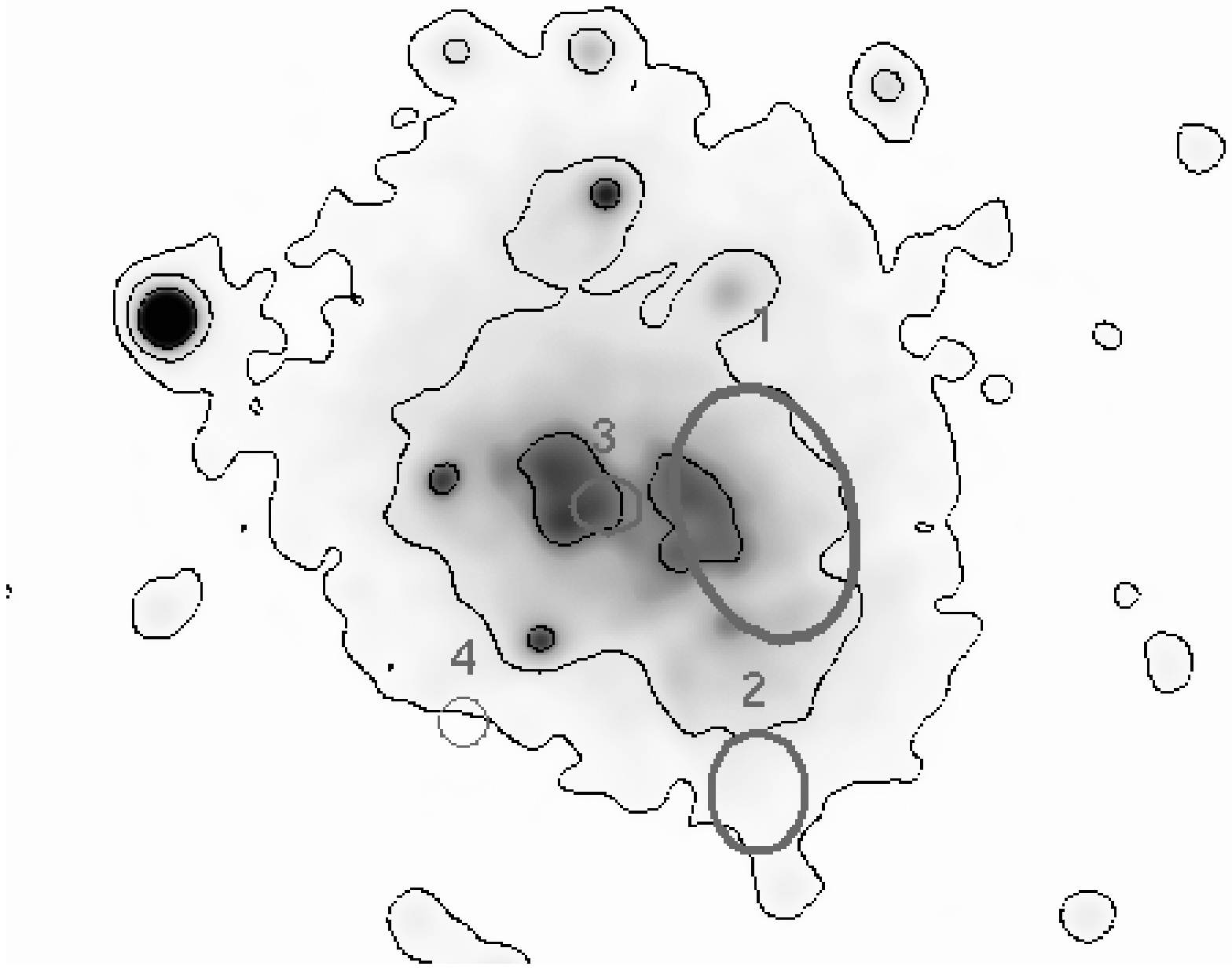}
\includegraphics[width=4.8cm]{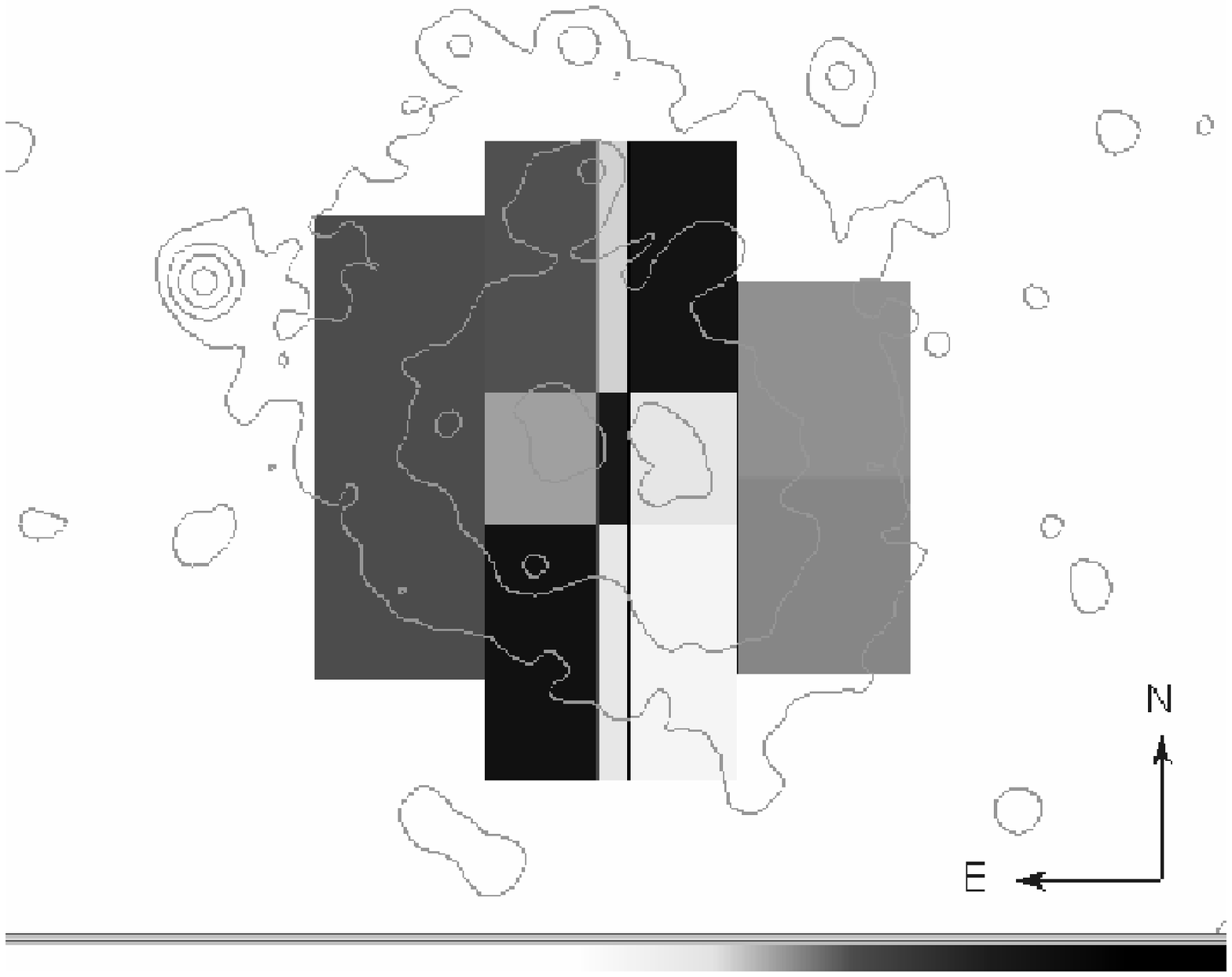}

\caption{Left: Adaptively smoothed image of the galaxy cluster CL~0939+4713. The core of CL~0939+4713 does not show a simple single central structure, but is composed by two subclumps. Right: Cluster temperature; superimposed are linearly spaced X-ray contours. A trend of higher temperatures is observed in the central region.}
\label{castillo_fig:fig2}
\end{figure}

\begin{figure}
\centering
\includegraphics[width=4.5cm]{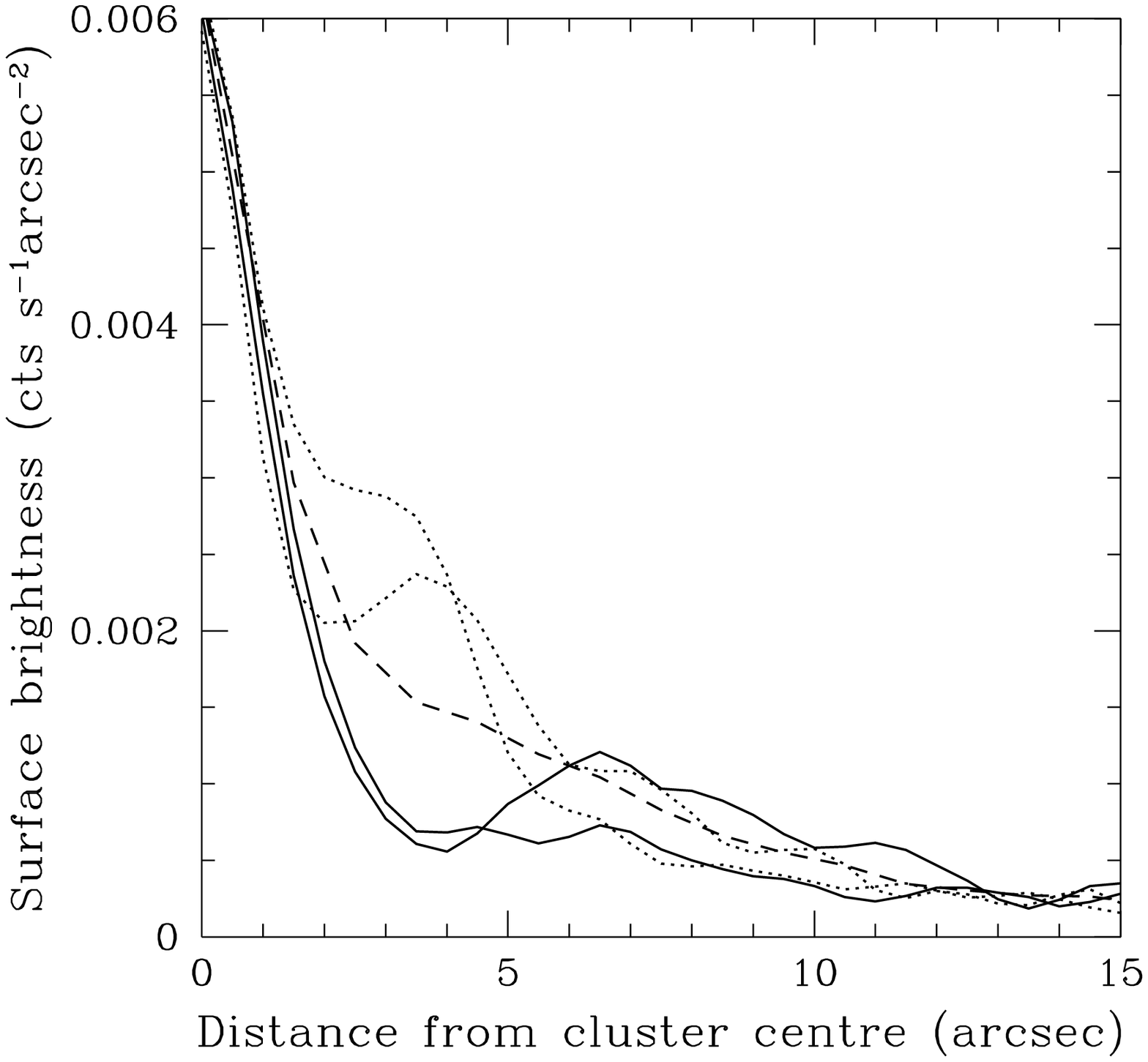}
\includegraphics[width=5.5cm]{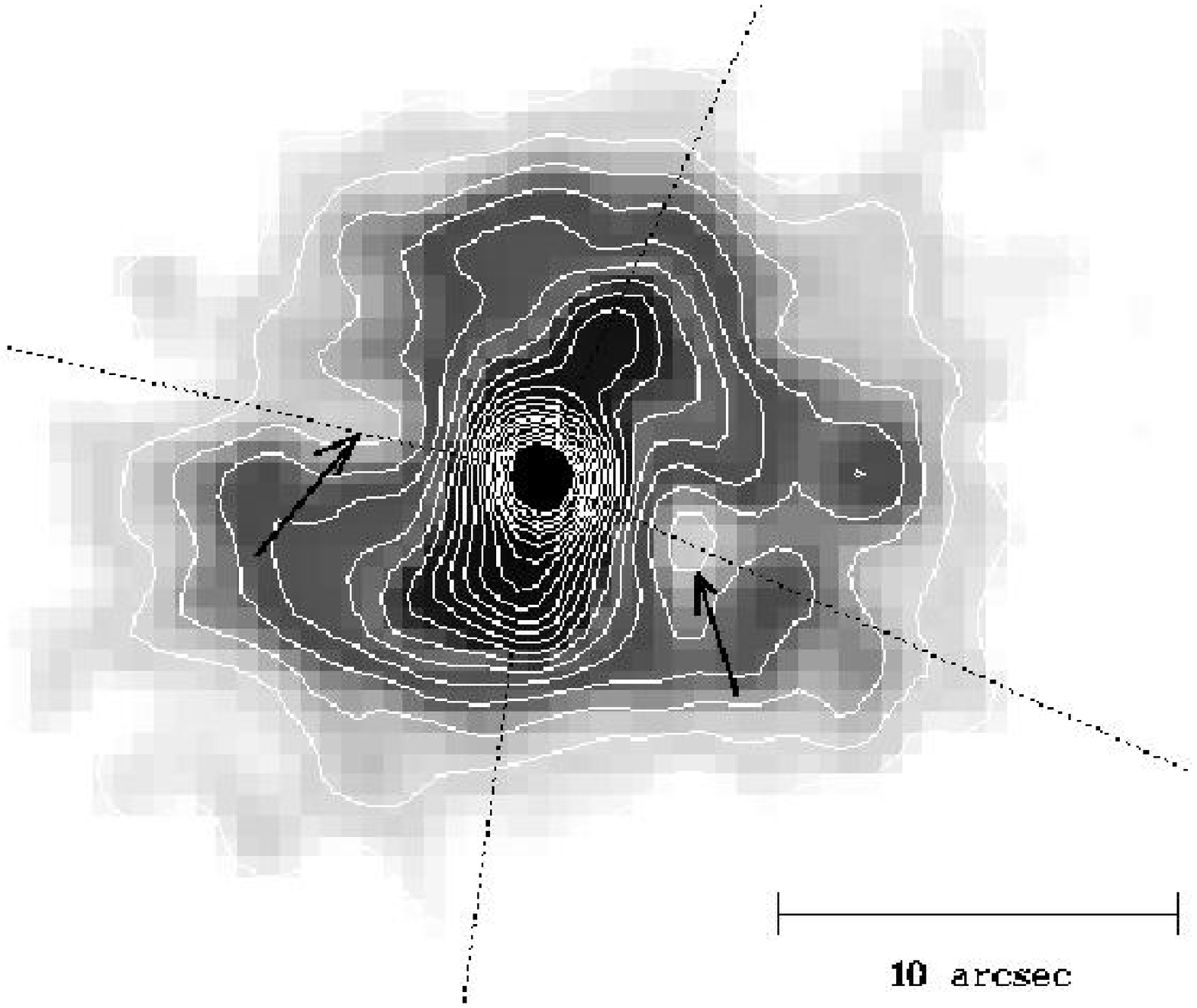}
\caption{Left: Surface brightness profiles in direction of the minima (solid lines), an in direction of the maxima (dotted lines). The dashed line shows an average profile integrated over all angles. Right: Smoothed image of the core of the cluster RBS797. The dotted lines show the directions along which the four surface brightness prifles were taken.}
\label{castillo_fig:fig3}
\end{figure}

We present the Chandra observation of the X-ray luminous, intermediate distant galaxy cluster RBS797 at $z=0.35$. In the central region the X-ray emission shows two pronounced X-ray minima, which are located opposite to each other with respect to the cluster centre. These depressiones suggest an interaction between the central radio galaxy and the intra-cluster medium, which would be the first detection in sucha distant cluster. The minima are symmetric relative to the cluster centre and very deep compared to similar features found in a few other nearby clusters (see Schindler {\it{et al.}} 2000 for more details).


\begin{chapthebibliography}{1}

\bibitem[]{} Cavaliere, A. \& Fusco-Femiano, R. 1976, A\&A, 49, 137
\bibitem[]{} Markevitch, M.L., Forman, W.R., Sarazin, C.L. \& Vikhlinin, A. 1998, ApJ, 503, 77
\bibitem[]{} Schindler, S. 1999, A\&A, 349, 435
\bibitem[]{} Schindler, S., Castillo-Morales, A., De Filippis, E., et al. 2000, A\&A, 376, L27
\bibitem[]{} Ponman, T.J., Cannon, D.B. \& Navarro, F.J. 1999, Nature, 397, 135



\end{chapthebibliography}
\end{document}